\documentclass[%
 reprint,
superscriptaddress,
nofootinbib,
 amsmath,amssymb,
 aps,
]{revtex4-1}
\usepackage{subcaption}
\usepackage{ragged2e}

\usepackage{graphicx}
\usepackage{dcolumn}
\usepackage{bm}

\usepackage{color}
\usepackage{hyperref}
\usepackage{cleveref}
\usepackage{comment}
\usepackage[normalem]{ulem}
\renewcommand{\l}{\left}
\renewcommand{\r}{\right}

\begin{abstract}
The existence of self-bound strange stars is a long-standing mystery in astrophysics. Future astrophysical data, even with improved precision, may not allow us to discriminate them from neutron stars, given the uncertainties in observational and theoretical modeling. In this work, we propose a unique strategy to distinguish strange stars from neutron stars using gravitational waves from binary compact star systems. We demonstrate that empirical relations connecting f-mode frequencies with tidal deformation are distinct for the two classes of compact objects, irrespective of their equations of state. Therefore simultaneous measurement of f-mode frequency and tidal deformability from  the inspiral phase of compact binary mergers with the next-generation detectors can provide smoking gun evidence for the presence of strange stars. This would have crucial implications not only in gravitational wave physics but multidisciplinary fields such as nuclear and high energy physics.
\end{abstract}

\begin{document}
\title{Prospects of identifying the presence of Strange Stars \\ using Gravitational Waves from binary systems}
\author{Bikram Keshari Pradhan}
\email{bikramp@iucaa.in}
\author{Swarnim Shirke}
\author{Debarati Chatterjee}

\affiliation{Inter-University Centre for Astronomy and Astrophysics, Pune,411007, India}
\maketitle

\section{Introduction}\label{sec:intro}
At the ultra-high densities in the interior of Neutron Stars (NS) exceeding supra-nuclear values \cite{Lattimer2004, Lattimer2021}, confined hadronic matter is expected to undergo a phase transition to deconfined quark matter~\cite{Collins1975, Glendenning1992, Alford2007, Baym2018}. Recent works suggested strong evidence for the presence of quark matter in NSs~\cite{Annala2020, Ferreira2020, Altiparmak2022, Ecker2022, Annala2022, Shirke2023, Annala2023, Han2023}. It is hypothesized that deconfined strange quark matter (SQM) could be the true ground state of matter in nature~\cite{Bodmer1971, Witten1984}. If true, NSs containing quark matter are expected to turn into exotic compact objects called Strange Stars (SS)~\cite{Alcock1986, Haensel1986}, which are self-bound objects entirely composed of three-flavored SQM containing up, down, and strange quarks.
The lack of first-principle calculations of the strong interaction
and the absence of terrestrial experiments on cold-dense matter makes the study of NSs and SSs essential. 
Their distinct internal composition and the possibility of pairing of quarks (color superconductivity) lead to a number of observational differences between SS and NS \cite{Alford2019}.
Detection of a SS would have extremely important implications for high-energy physics and astrophysics. Such a detection would pin down the transition density, reveal the true ground state of matter, constrain microphysics, and, at the same time, establish a new class of astrophysical objects.


 
\begin{figure}
    \centering
    \includegraphics[width=\linewidth]{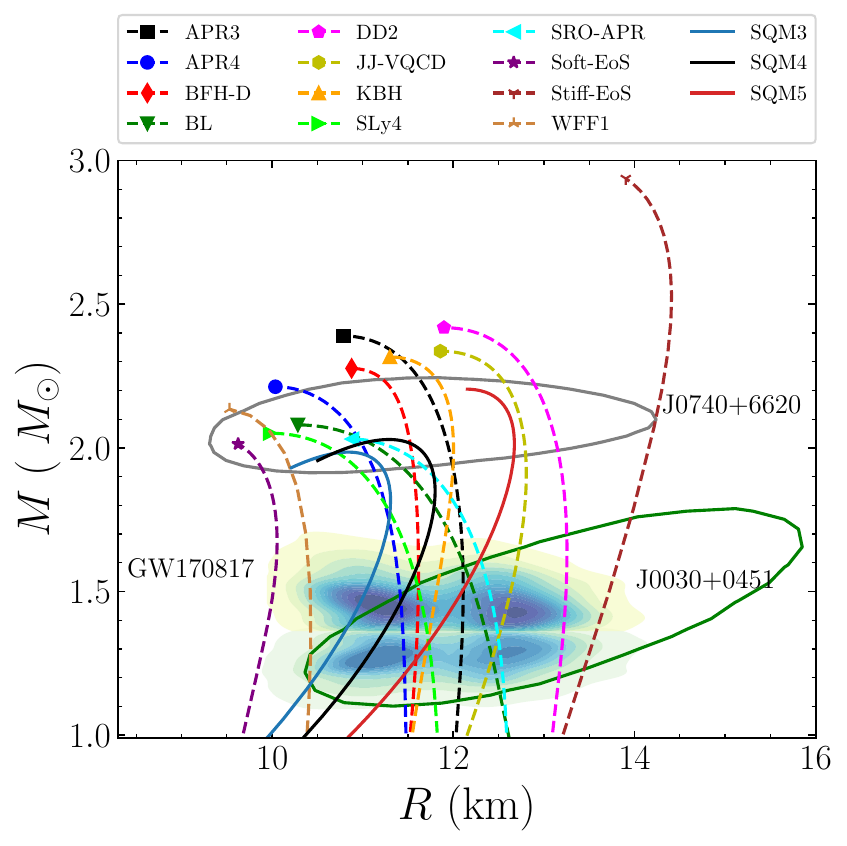}
    \caption{ \justifying The mass-radius relation for compact stars with selected realistic EOSs for Neutron stars, Hybrid stars, and Strange quark stars.  NS and HS EOS models:  APR4, APR3~\cite{APR,Douchin2001}, SLy4~\cite{Gulminelli2015}, BL~\cite{BL2018}, DD2~\cite{CHABANAT1998231}, SRO-APR~\cite{SROAPR}, WFF1~\cite{WFF1}, Soft and Stiff EoS from~\cite{Hebeler_2013}, BFH-D~\cite{Baym_2019,TOGASHI2017}, KBH ($\rm QHC21-AT$)~\cite{KBH_2022}, JJ-VQCD~\cite{JJVQCD}. In addition to the NS and HS EOS models, we have also displayed a few strange star EOS models satisfying all astrophysical observations, such as the SQM3 model from~\cite{Lattimer2001}. We have also considered two strange star EOS models described by a phenomenological model based on MIT bag model~\cite{Alford2005, Alford2012a, Alford2012b}: (1) represented by SQM4 with bag parameter ($B^{1/4}=138$ MeV), strange quark mass ($m_s=150$ MeV ), interaction strength parameter ($a_4=0.7$) and (2) represented by  SQM5 with bag parameter ($B^{1/4}=135$ MeV), strange quark mass ($m_s=100$ MeV), interaction strength parameter ($a_4=1$). The uncertainties on the $M-R$ measurements (90\% contour levels) for  PSR J0740+6620 ~\cite{riley2021} and PSR J0030+0451 ~\cite{Riley_2019} have also shown. The mass-radius estimates of the two companion neutron stars of the merger event GW170817 are shown by the shaded area labeled GW170817. }
    \label{fig:MR}
\end{figure}

 There have been extensive searches for signatures of SQM~\cite{Weber2005}. Although there have been a few candidates for SSs in the past, they either have been ruled out, or there is no conclusive evidence for their existence yet~\cite{Weber2005}. Hence, their existence is still an open question. The recent observation of the compact and lightest supernova remnant HESS J1731-347\cite{Doroshenko2022} has become a favorite candidate for SS due to its low mass and radius~\cite{DiClemente2022, Horvath}; however, SSs still remain degenerate with NSs in high mass range (see~\cref{fig:MR}), making them indistinguishable even with future precise mass-radius measurements. Though strange star sequences differ from NS models and can be distinct at very low masses ($M<1M_{\odot}$), strange star configurations in the range $M\geq 1 M_{\odot}$ show significant degeneracy with the NS (or HS) configuration, makeing their distinguishability more challenging. Several theoretical efforts have been made to discriminate the presence of strange quark stars from neutron stars by studying the sequential behavior of stellar properties such as radius (R), tidal deformability ($\Lambda$) or f-mode characteristics (frequency and damping time) as a function of mass \cite{YipChuLeung1999, Kojima2002, Sotani2003, Sotani2004}. Though a large number of very precise observations at different mass regimes can reveal the true nature of the compact star sequences, the expected observational uncertainties and the lack of understanding of the enigmatic high-density matter may not allow us to distinguish strange stars from neutron stars. 
 

In this work, we aim to distinguish the presence of strange stars from neutron stars, particularly those with masses $\geq 1M_{\odot}$, using gravitational wave observations from binary systems in their inspiral phase. As discussed in~\cite{Wen2019}, strange stars do not adhere to the same Universal Relations (UR) as neutron stars, particularly the f-Love relation, which connects the mass-scaled quadrupolar ($\ell=2$) f-mode angular frequency  $\bar{\omega}_2=M\omega_2$ with the tidal parameter $\Lambda$ (see,~\cref{app:f_love_relation})~\cite{Chan2014}. GW observations offer a unique opportunity to independently measure the f-mode frequency and tidal deformability simultaneously~\cite{Williams2022,Gamba2022}. Hence, it can be used to probe the nature of the compact objects using the plane of  $\bar{\omega}_2$ and $\Lambda$ (see~\cref{sec:method} for details). This article introduces a novel methodology, reliant upon the measurements $\bar{\omega}_2$ and $\Lambda$, that serves as a discerning tool for identifying the presence of strange stars in binary systems. While the detection of the binary neutron star event GW170817~\cite{AbbottAJL848,AbbottPRL119,AbbottPRL121,AbbottPRX} was a significant step in multi-messenger astronomy, there are no significant observed constraints on  f-mode characteristics of the companions ~\cite{Pratten2020,Williams2022,Gamba2022,Wen2019}. Our analysis can become accessible once the next generation GW detectors act to operate, as it involves the measurement of f-mode characteristics.




\begin{figure*}
    \centering
    \includegraphics[width=\linewidth]{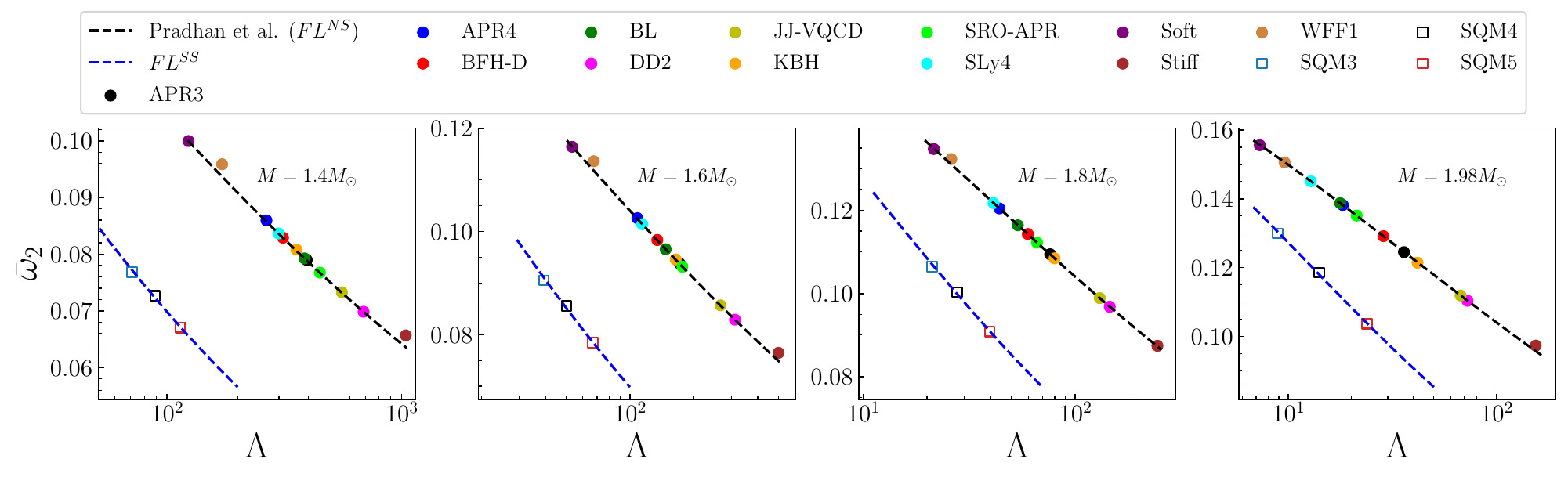}
    \caption{ \justifying The dimensionless f-mode angular frequency $\bar{\omega}_2$ as a function of $\Lambda$ for a different  fixed mass configurations for different EOSs.}
    \label{fig:f_love_theory}
\end{figure*}
\section{Formalism}\label{sec:method}
Strange stars exhibit a distinct f-Love relation compared to NSs in every range of $\Lambda$, emphasizing their unique characteristics (see,~\cref{app:f_love_relation}). This distinction implies that by measuring the parameters $\Lambda$ and the dimensionless quantity $\bar{\omega}_2=M\omega_2$ of compact stars, regardless of their mass, one can effectively differentiate SSs from NSs. For a better understanding of this concept, in~\cref{fig:f_love_theory}, we have presented the f-Love relations for NSs and SSs alongside the numerically derived values based on representative Equation of State (EOS) models as shown in~\cref{fig:MR}. The f-Love relation for NSs ($FL^{NS}$) has been adapted from~\cite{Pradhan2023}, while for the f-Love relation corresponding to SS ($FL^{SS}$), it has been established based on a comprehensive analysis of approximately $\sim 10^4$ SS configurations. The f-Love relations are given in~\cref{app:f_love_relation}. Evidently, the f-Love relation effectively distinguishes between these two distinct families of compact objects, NS  and SS. The deviation in the f-Love relation across all mass ranges indicates that analyzing the favorable f-Love relation (among the NS and SS) using the joint measurement of the tidal deformability ($\Lambda$) and the f-mode parameter ($\bar{\omega}_2$) in binary companions can differentiate their nature, breaking ambiguity and potentially confirming the presence of strange stars in binary systems (and hence in nature)  otherwise challenging even with the very precise measurement of the other stellar properties.
The existence of two distinct f-Love relations for different families of compact objects suggests that simultaneous measurements of $\Lambda$ and $\bar{\omega}_2$ can be utilized for statistical tests, such as Bayes factor comparison or Odds ratio methodology. These tests can help unveil the preferred f-Love relations among $Fl^{NS}$ and $FL^{SS}$, offering insights into the nature of compact objects.

 In the following section, we will simulate the binary GW signal and perform a Bayesian parameter estimation of the simulated events. We will then examine the Odds ratio or Bayes factor to statistically determine the preferred f-Love relation and hence the nature of the compact object. It has been speculated and discussed that f-mode characteristics could be measured accurately with the next-generation GW detectors~\cite{Williams2022,Pratten2021,Gamba2022}. Measurement of f-mode parameters being the key requirement of our analysis, we consider the next-generation Cosmic Explorer (CE)~\cite{CE,CE2} with the proposed design sensitivity~\footnote{\url{https://dcc.ligo.org/LIGO-P1600143/public}} and placed at the location of Hanford (H1) and Livingstone (L1) of the current LIGO detectors. We simulate the binary GW signal with the frequency domain TaylorF2 waveform model, including  3.5 PN (Post Newtonian) point particle phase, adiabatic tidal effects up to 7.5PN order, and the ready-to-use quadrupolar f-mode dynamical tidal correction from~\cite{Schmidt2019}. The simulated waveform starts at a minimum frequency of 10Hz and truncates at a frequency that is the minimum of the contact frequency~\cite{Agathos2015} and the frequency of the innermost circular orbit.
  
  To measure the f-mode parameters, we perform the Bayesian parameter estimation of the simulated events using the nested sampling $\textit{dynesty}$~\cite{Dynesty} as implemented in the python package  \textit{bilby\_pipe}~\cite{Bilby_2019}. One of our motivations is distinguishing the presence of strange stars, in particular the overlap region, say $M\in [1,2]M_{\odot}$. Accordingly, we keep the injections in this mass region. We independently measure the f-mode parameters and the tidal deformability and avoid the use of any universal relations in this stage~\footnote{In contrast, in several works the NS f-Love relation has been used to investigate the effect of the f-mode dynamical tides~\cite{Pratten2021,Pradhan2023,Gamba2022}.}.  
  
  Given an event with observed data $d$, for the two different f-Love relations: (1) the f-Love relation from~\cite{Pradhan2023} for NSs (will represent this as $FL^{NS}$) and (2) f-Love relation for strange stars (will represent this as $FL^{SS}$), the Odds ratio can be defined as,
    \begin{equation}\label{eqn:odds}
    \mathcal{O}^{NS}_{SS}=\frac{P(FL^{NS}|d)}{P(FL^{SS}|d)}
    \end{equation}
where $P(FL^{i}|d)$ is the  probability of the f-Love relation ($FL^i$) for the data `d'. Now, using Bayes theorem,
\begin{equation}\label{eqn:odds2}
    \mathcal{O}^{NS}_{SS}=\frac{P(d|FL^{NS})}{P(d|FL^{SS})} \times \frac{\pi(FL^{NS})}{\pi(FL^{SS})}
\end{equation}

 \begin{align}\label{eqn:liklhd}
        P(d|FL^i)
              &\propto  \int d\Lambda  P(d|\Lambda,\bar{\omega}_2=FL^i(\Lambda))\  P(\Lambda|FL)  &
     \end{align}
The quantity $P(d|\Lambda,\bar{\omega}_2$) is obtained by  dividing the priors $\pi_{\rm PE}({\Lambda,\bar{\omega}_2})$ from the corresponding posteriors $P({\Lambda,\bar{\omega}_2} \mid d)$ while performing the  parameters estimation using $bilby-pipe$. As our analysis depends upon the joint distribution of ($\Lambda,\bar{\omega}_2$), we marginalize over all other parameters (e.g., the component masses). Furthermore, we keep the $P(\Lambda|FL)$ to be uniform distribution $\in [1,5000]$ while calculating~\cref{eqn:liklhd}.

Further considering both the f-Love relations as equally likely, i.e., $\pi(FL^{NS})=\pi(FL^{SS})$, the odd's ratio can be reduced to the Bayes factor comparison. 
\begin{align}
    \mathcal{O}^{NS}_{SS}=\mathcal{B}_{SS}^{NS} \times \frac{\pi(FL^{NS})}{\pi(FL^{SS})} \nonumber 
\end{align}
where, 
\begin{equation}\label{eq:Bayesfactor}
     \mathcal{B}_{SS}^{NS}=\frac{P(d|FL^{NS})}{P(d|FL^{SS})} 
\end{equation}

Hence, by examining the Bayes factor, one can arrive at a statistical decision concerning the nature of the companion(s) in binary systems. For instance, if $\log_{10}[{\mathcal{B}_{SS}^{NS}}]$ is significantly negative, the observation favors the strange star f-Love relation over the neutron star f-Love relation, pointing toward the presence of a strange star. Notably, for the expressions in~\cref{eqn:odds,eqn:odds2,eqn:liklhd,eq:Bayesfactor} to yield meaningful results, it is essential to independently measure the posterior distributions of ($\Lambda, \bar{\omega}_2$) for each companion. This signifies that the methodology can be applied individually to each companion to ascertain if one or both of them are strange stars, thereby indicating the potential existence of hybrid binary systems comprising both neutron stars and strange stars.

Another interesting aspect of this approach is that it can confirm the presence of strange stars without requiring a series of precise measurements, such as examining sequences of stellar observables like ($M-R$) or ($M-\Lambda$). In fact, this methodology can be effectively applied to each companion in a single binary system detection. Consequently, a single reliable detection of GW from a binary system with f-mode measurements has the potential to serve as compelling evidence for the presence of strange stars (if they are indeed present).

 \section{Results}~\label{sec:result}

For the demonstration of our methodology, we consider a  GW170817 mass configuration-like binary system at a luminosity distance $D_L=50$ Mpc with the source frame component masses as $M_1=1.475M_{\odot}$ and $M_2=1.26M_{\odot}$.  Assuming f-modes can be observed with the next-generation GW detectors, we consider the detector configuration mentioned in~\cref{sec:method}. Assuming the companions are strange stars, we assign the other required properties such as $\Lambda_i$, $\bar{\omega}_i$ of the companions of masses $M_i$  corresponding to the SQM3 EOS model for simulated GW signal. During the parameter estimation with \textit{bilby\_pipe}, we sample with the uniform priors on the component masses ($m_i\in [0.5,3.]$), uniform priors on individual tidal deformability $\Lambda_i \in [1,5000]$ and uniform priors on the $\bar{\omega}_{2,i} \in [0.025,0.18]$. 

\begin{figure}
     \centering
     \includegraphics[width=\linewidth]{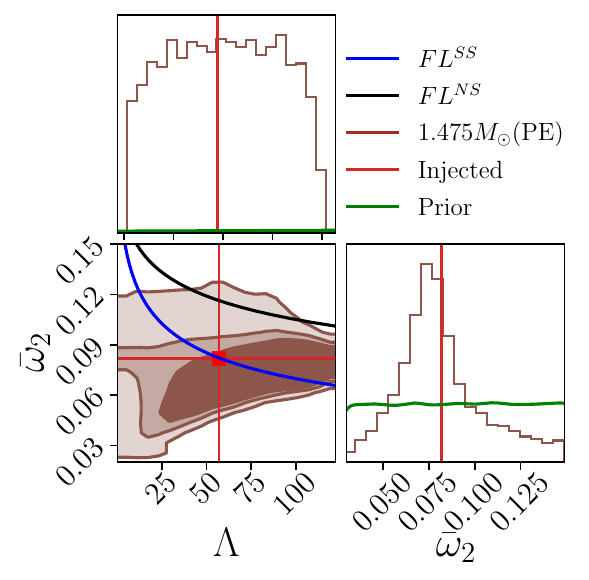}
     \caption{ \justifying Joint and marginalized posterior distribution of the recovered parameters $\Lambda$ and $\bar{\omega}_2$, corresponding to the primary component $M_1=1.475M_{\odot}$ of the considered event in~\cref{sec:result}. The injected values and the different f-Love relations are also displayed.}
     \label{fig:f_lambda_posterior}
 \end{figure}

\begin{figure}
     \centering
     \includegraphics[width=\linewidth]{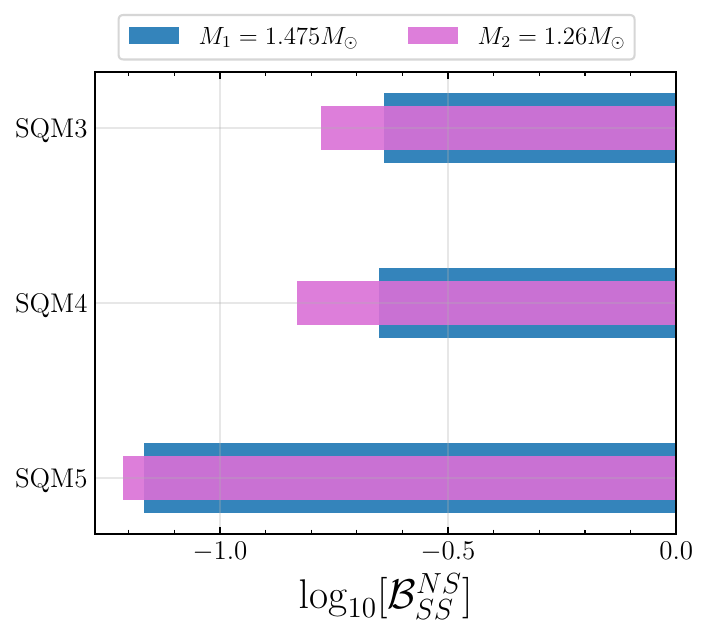}
     \caption{ \justifying Bar plot showing the value of $\log_{10}{\mathcal{B}^{NS}_{SS}}$ for different assumed SS EOSs,  obtained following~\cref{sec:method}, for the companions of the event considered in~\cref{sec:result}, with masses $M_1=1.475M_{\odot}$ (blue) and $M_2=1.26M_{\odot}$ (magenta).}
     \label{fig:Bayes_factor_SQM}
 \end{figure}

For the considered event, the posterior distribution of recovered ($\Lambda,\ \bar{\omega}_2$) for the primary component with mass  $M_1=1.475M_{\odot}$ corresponding to SQM3 EOS is displayed in~\cref{fig:f_lambda_posterior}. The two f-Love relations representing the NS f-Love relation ($FL^{NS}$) and SS f-Love relation ($FL^{SS}$) are also shown in~\cref{fig:f_lambda_posterior}. Although looking at~\cref{fig:f_lambda_posterior}, one can conclude that the  Strange star f-Love relation seems more likely to explain the data compared to $FL^{SS}$, there is still substantial overlap with the $FL^{NS}$. For the statistical evidence, we obtain the Bayes factor for each of the companions as given in~\cref{eq:Bayesfactor} of~\cref{sec:method} and displayed in the~\cref{fig:Bayes_factor_SQM}. Following the interpretation from~\cite{Kass1995}, one can divide the Bayes factor analysis to (a) $\log_{10} {\mathcal{B}^{NS}_{SS}}\leq -2$, implies decisive evidence in favor of strange star over NS (b) $-2 < \log_{10} {\mathcal{B}^{NS}_{SS}}\leq -1$, implies strong evidence in favor of strange star over NS (c) $-1 < \log_{10} {\mathcal{B}^{NS}_{SS}}< -1/2$, implies substantial evidence in favor of strange star over NS and, (d) $\log_{10} {\mathcal{B}^{NS}_{SS}}\geq -1/2$, implies insubstantial evidence in favor of strange star over NS. From~\cref{fig:Bayes_factor_SQM}, one can conclude that there is substantial evidence in favor of companions as SSs, or in other words, our methodology results in substantial evidence supporting the companion as a strange star and leads to their distinguishability. 

We test the sensitivity of the proposed methodology to the stiffness of the EOS  model by considering two additional strange star EOS models, SQM4 and SQM5, for the injection GW. We then repeat our methodology to calculate the Bayes factor ($\mathcal{B}^{NS}_{SS}$) and display the resulting Bayes factor in~\cref{fig:Bayes_factor_SQM}. From~\cref{fig:Bayes_factor_SQM}, one can conclude that irrespective of the stiffness of the assumed EOS, $\mathcal{B}^{NS}_{SS}$  always indicates substantial evidence in favor that the companions are strange stars. Interestingly, with the stiffer assumed SQM5 model, we find strong evidence supporting that the companions are strange stars. Regardless of the chosen EOS models, our methodology consistently indicates enhanced distinguishability for strange stars, particularly those with lower masses (but still $M \geq 1M_{\odot}$), where the properties of SSs exhibit degeneracy with those of NSs.). The better prediction at the lower mass regime can be explained by looking at~\cref{fig:f_love_theory}, which indicates that the separation among the two different f-Love relations increases with decreasing the mass. In addition,  due to the tight nature of f-Love relations, our conclusions remain unaltered in considering the uncertainties on the f-Love relation in our analysis.

Furthermore, we have tested our methodology for binary systems with NS, assuming a wide range of NS EOS models for the injections. For the different NS EOS models considered, we find that $\log_{10}{\mathcal{B}^{NS}_{SS}} >0.5$, confirming 
that the companion(s) is (are) an NS (see~\cref{app:NS_EOS} ). It is worth mentioning that we can always reveal the assumed nature of the companions using the simultaneous measurement of $\Lambda$ and $\bar{\omega}_2$. Hence, for a real observation of GW from a binary system,  our methodology can predict the true nature of the binary companions, whether they are strange stars or neutron stars, without any prior information about the companions or about the EOS model.


\section{Discussion}
In this article, we demonstrate a novel methodology to identify the presence of strange stars utilizing the GW wave observation from binary systems. We find that, with the simultaneous measurement of f-mode frequency and the tidal deformability by next-generation GW detectors, strange stars can be distinguished from NSs, and their presence can be confirmed. Though we do not explore all the uncertainties in the SS EOS sector, we have tested our methodology for EOSs with different stiffness and with different mass ranges, where the SS properties are degenerate with those of NS and predict their distinguishability. However, apart from this, since the f-love relations for NS and SS are universal, our analysis is independent of EOS models. The formalism developed can also be applied to individual companions of the binary and does not require a series of detections to confirm the presence of SSs. This work can be further improved by considering the inclusion of spin and eccentricity as they can affect 
the excitation of f-mode in binary~\cite{Ma2020,Kuan2022,Chirenti_2017,Steinhoff2021,Pnigouras2022} and subject to future investigation. The methodology can be further checked with   GW waveform models, allowing the measurement of f-mode frequency other than the TaylorF2 waveform model used in this work. 

As we usher into a new era of multi-messenger astronomy since the GW170817 event~\cite{AbbottAJL848,AbbottPRL119}, it has been well-established that GW signals from NS binary systems can constrain the microscopic EOS~\cite{Annala2018,AbbottPRL121,AbbottPRX}. Our work shows how, apart from this, the GW signals, solely from the inspiral phase, can also be used to comment conclusively on the longstanding problem of the existence of SSs. This will settle the SQM hypothesis by establishing the nature of the true ground state of matter and reveal the hadron-quark phase transition density. This adds to how astrophysics and gravitational wave physics will be crucial in discerning the properties of strongly interacting matter in the near future. For this reason, the current work is relevant to a wide range of physics disciplines, including not only astrophysics and gravitational waves physics but also high-energy particle physics and nuclear physics. Subsequent detections and observations of SSs can impose stringent constraints on phenomenological models for cold quark matter. The results presented here are timely, with the next-generation GW detectors already underway.

\section{Acknowledgments}
The authors gratefully acknowledge the use of the Sarathi cluster at IUCAA accessed through the LIGO-Virgo-KAGRA Collaboration. B.K.P. acknowledges the  IUCAA HPC computing facility Pegasus for the computational/numerical work.
\appendix
\section{f-Love relations for NS and SSs}\label{app:f_love_relation}
The universality among the mass-scaled f-mode angular frequency and tidal deformability parameter  ($\Lambda$)  (or love number) was first introduced in ~\cite{Chan2014} and later updated in subsequent works in the literature. In this work, we use the f-Love relation for the NSs developed in our previous work~\cite{Pradhan2023}, where we have considered a wide range of EOSs. As our methodology resides in the model selection for the URs, we first develop the URs for the strange star family. The EOS of quark matter at stellar densities is not known from first-principle calculations. To describe the quark matter EOS, we use the quartic parametrization of the phenomenological MIT bag model \cite{Farhi1984} as described in \cite{Alford2005, Alford2012a, Alford2012b}. The ungapped quark matter that we consider has three model parameters, namely $a_4$, $B$, and $m_s$, accounting for the strong interaction strength, the bag constant, and the strange quark mass, respectively.
The values of $a_4$ and $B$ are chosen such that SSs are produced under the SQM hypothesis.

\begin{figure}[ht]
    \includegraphics[width=\linewidth]{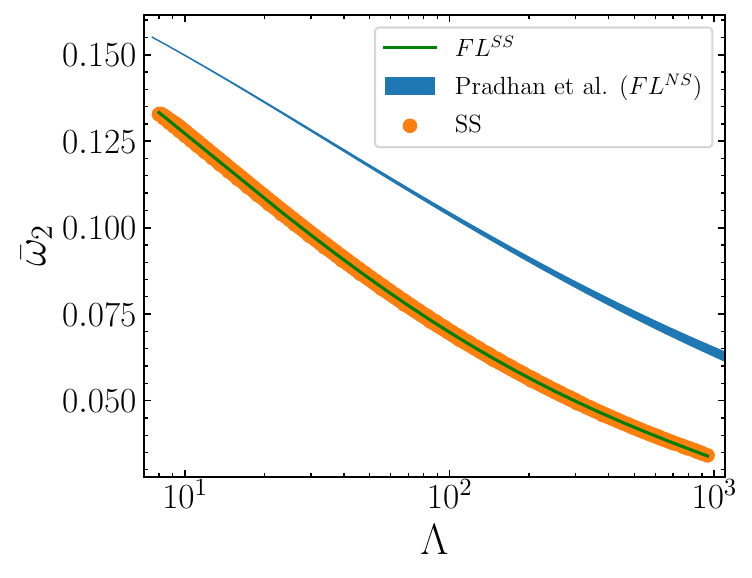}
    \caption{The universal relation (f-Love relation) among the f-mode parameter $\bar{\omega}_2$ and the dimensionless tidal deformability $\Lambda$. The f-Love relation shown for the NSs is adapted from ~\cite{Pradhan2023} along with the uncertainty displayed in blue in color (represented as $FL^{NS}$). In contrast to the NSs, the strange star f-Love relation is obtained by considering $\sim 10^4$ SS configurations.}
    \label{fig:f_love_relation}
\end{figure}

The leading-order dynamical tidal correction, which is also detectable, is the quadrupolar correction. Therefore, in our discussion, we focus on the universal relation between the quadrupolar f-mode frequency ($\omega_2$) and the quadrupolar tidal deformability parameter ($\Lambda$). We express this relation as a polynomial fit, which we refer to as the f-Love relation. To obtain the f-Love relation, we solved the f-mode for $\sim 10^4$ SSs and displayed in~\cref{fig:f_love_relation}. For comparison, we have included the f-Love relation for NSs from our previous study~\cite{Pradhan2023}. The fit parameters for the f-Love relation of strange stars ($FL^{SS}$) along with the relation from~\cite{Pradhan2023} are given in~\cref{tab:f_Love_fitparameters}.

\begin{eqnarray}
    m\omega_2=\bar{\omega}_2=\sum_{k=0}^{6} a_k \l[\ln{(\Lambda_2)}\r]^k, \label{eqn:tidal_fit}
\end{eqnarray}

\begin{table*}[ht]
    \centering
    \begin{tabular}{c |  c  c c c c c c}
    \hline \hline
      Relation&$a_0$   & $a_1$ & $a_2$ &$a_3$ &$a_4$ &$a_5$ &$a_6$ \\
    \hline
     $FL^{NS}$~\cite{Pradhan2023} &1.820$\times 10^{-1}$& -6.665$\times 10^{-3}$& -4.212$\times 10^{-3}$&  4.724$\times 10^{-4}$&
        -1.030$\times 10^{-6}$& -2.139$\times 10^{-6}$&  8.763$\times 10^{-8}$\\
         $FL^{SS}$ [This Work]&1.61$\times 10^{-1}$& -1.385$\times 10^{-2}$& -2.235$\times 10^{-2}$&  5.881$\times 10^{-3}$&
        -8.007$\times 10^{-4}$& -5.929$\times 10^{-5}$&  -1.88$\times 10^{-6}$\\
     \hline \hline
    \end{tabular}
    \caption{Values of the fit parameters for f-Love relation~\eqref{eqn:tidal_fit} for NSs ($FL^{NS}$) and SSs ($FL^{SS}$).}
    \label{tab:f_Love_fitparameters}
\end{table*}

\section{Analyzing the $\log_{10}[\mathcal{B}_{SS}^{NS}]$ for the NS EOSs.}~\label{app:NS_EOS}
\begin{figure}[ht]
    \centering
    \includegraphics[width=\linewidth]{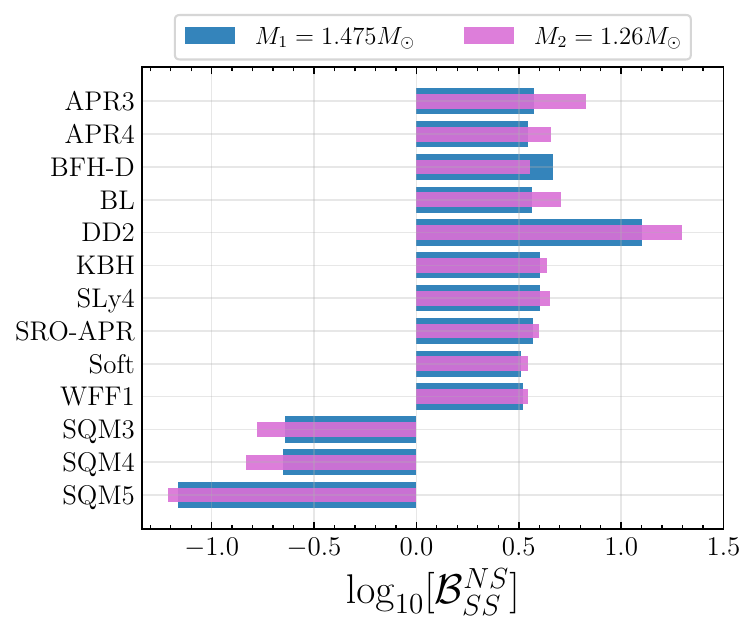}
    \caption{Resulting $\log_{10}[\mathcal{B}_{SS}^{NS}]$ for the companions of a binary system with masses $M_1=1.475M{\odot}$ and $M_1=1.26M{\odot}$ under the assumption of different EOS model for simulated events. For comparison, results for both the NS EOS and SS EOSs are displayed in a single plot. }
    \label{fig:Bayes_factor_all}
\end{figure}

 Though our article primarily focuses on the methodology for identifying SSs, we also extended our methodology to test it with NSs and discuss the results here. In our test scenario, we assumed a binary system consisting of NSs following generic NS EOSs. We repeated the methodology described in~\cref{sec:method}, and the results are presented in~\cref{fig:Bayes_factor_all}. In~\cref{fig:Bayes_factor_all}, we display the $\log_{10}[\mathcal{B}_{SS}^{NS}]$, for various NS EOSs, as well as for SS EOSs, for better comparison. Notably, regardless of the choice of hadronic or hybrid NS EOS, $\log_{10}[\mathcal{B}_{SS}^{NS}]$ always turns out to be $\sim 0.5$. This trend favors the NS f-Love relation over $FL^{SS}$ and suggests that the companions in these binary systems are likely to be NSs, supporting our initial assumption. As a result, we conclude that the independent measurement of the tidal deformability parameter ($\Lambda$) and the f-mode frequency ($\bar{\omega}_2$) of a compact star from a gravitational wave (GW) event, combined with the application of our methodology, can distinguish the nature of the compact object, whether it is a strange star or a neutron star.

\bibliography{Pradhan}
\end{document}